\documentclass[useAMS,usegraphicx]{mn2e}
\usepackage{rotate}
\usepackage{times}
\newif\ifAMStwofonts
\AMStwofontstrue
\def\gsim{\mathrel{\hbox{\rlap{\hbox{\lower4pt\hbox{$\sim$}}}\hbox{$>$}}}}
\def\lsim{\mathrel{\hbox{\rlap{\hbox{\lower4pt\hbox{$\sim$}}}\hbox{$<$}}}}

\def\Msun{M$_{\odot}$}

\def\xmm{{\it XMM-Newton}}

\def\asca{{\it ASCA}}
\def\sax{{\it BeppoSAX}}

\def\xmm{{\it XMM-Newton}}

\def\et{{et al.\ }}
\def\mcg{{MCG--6-30-15}}

\def\Msun{\hbox{$\rm\thinspace M_{\odot}$}}

\title[The iron line in \mcg\ from \xmm ]
      {The iron line in \mcg\ from \xmm: evidence for gravitational
light-bending?}

\author[Fabian \et]
       {A. C. Fabian\thanks{E-mail: acf@ast.cam.ac.uk}
        and S. Vaughan \\
Institute of Astronomy, University of Cambridge, Madingley Road, Cambridge CB3 0HA\\
}
\date{Accepted: 29/1/2003}
\pagerange{\pageref{firstpage}--\pageref{lastpage}}
\pubyear{2002}
\begin{document}
\maketitle
\label{firstpage}

\begin{abstract}
The lack of variability of the broad iron line seen in the Seyfert 1
galaxy \mcg\ is studied using the EPIC pn data obtained from a
2001 observation of 325~ks, during which the count rate from the
source varied by a factor of 5. The spectrum of 80~ks of data from
when the source was bright, using the the lowest 10~ks as background,
is well fitted over the 3--10~keV band with a simple power-law of
photon index $\Gamma=2.11$. The source spectrum can therefore be
decomposed into an approximately constant component, containing a
strong iron emission line, and a variable power law component.
Assuming that the power-law continuum extends down to 0.5~keV enables
us to deduce the absorption acting on the nucleus. A simple model
involving Galactic absorption and 2 photoelectric edges (at 0.72 and
0.86~keV) is adequate for the EPIC spectrum above 1~keV. It still
gives a fair representation of the spectrum at lower energies where
absorption lines, UTAs etc are expected. Applying the absorption
model to the low flux spectrum enables us to characterize it as
reflection-dominated with iron about 3 times Solar.  Fitting this
model, with an additional power-law continuum, to $32\times 10$~ks
spectra of the whole observation reveals that $\Gamma$ lies mostly
between 2 and 2.3 and the normalization of the reflection-dominated
component varies by up to 25 per cent. The breadth of the iron line
suggests that most of the illumination originates from about 2
gravitational radii. Gravitational light bending will be very strong
there, causing the observed continuum to be a strong function of the
source height and much of the continuum radiation to return to the
disk. Together these effects can explain the otherwise puzzling
disconnectedness of the continuum and reflection components of
\mcg.
\end{abstract}

\begin{keywords}
galaxies: active -- galaxies: Seyfert: general -- galaxies:
individual: \mcg\ -- X-ray: galaxies 
\end{keywords}

\section{Introduction}

The X-ray spectrum of the bright Seyfert 1 galaxy \mcg\
($z=0.00775$) has a broad emission feature stretching from below 4~keV
to about 7~keV. The shape of this feature, first clearly resolved with
\asca\ by Tanaka \et (1995), is skewed and has a peak at about
6.4~keV. This profile is consistent with that predicted from iron
fluorescence from an accretion disc inclined at 30~deg and extending
down to within about 6 gravitational radii ($6r_{\rm g} = 6GM/c^2$) of
a black hole (Fabian \et 1989; Laor 1991). In part of the
\asca\ observation the line extended below 4~keV (Iwasawa \et 1996)
which means that the emission originates at radii less than $6r_{\rm
g}$, possibly due to the black hole spinning. \xmm\ has now observed
\mcg\ twice (Wilms \et 2001; Fabian \et 2002a) and in both cases the
line extended down to about 3~keV.

The X-ray continuum emission of \mcg\ is highly variable (see Vaughan,
Fabian \& Nandra 2002 for a recent analysis). If the observed
continuum drives the iron fluorescence then the line flux should
respond to variations in the incident continuum on timescales
comparable to the light-crossing, or hydrodynamical time of the inner
accretion disc (Fabian et al 1989; Stella 1990; Matt \& Perola 1992;
Reynolds et al 1999). This timescale ($\sim 100M_6$~s for reflection
from within $10r_{\rm g}$ around a black hole of mass $10^6
M_6$~\Msun) is short enough that a single, long observation can span
many light-crossing times. This has motivated observational efforts to
find variations in the line flux (e.g. Iwasawa \et 1996, 1999;
Reynolds 2000; Vaughan \& Edelson 2001; Shih, Iwasawa \& Fabian 2002).
These analyses indicated that the iron line in \mcg\ was indeed
variable on timescales of $\sim 10^4$~s, but that the amplitude of the
variations was considerably less than expected and the variations were
not correlated with the observed continuum. The data also showed a
strong correlation between the photon index of the best-fitting
power-law and the flux of the source.

In this letter we examine the later, long \xmm\ observation in more
detail using a simple two-component model to explain the
observed spectral variability. Specifically, the data are compared to
the model proposed by Shih \et (2002) comprising  a
highly variable power-law component plus a constant (or less variable)
harder component carrying the iron line. This gives a good fit to the
data, with the harder, line-carrying component dominating lowest flux
states of the observation. On the assumption that the variable
component of the spectrum is just a power-law, the variability enables
us to determine the attenuation at low energies due to both Galactic
absorption and the warm absorber in \mcg. It also enables us to
demonstrate that there is no subtle additional absorption influencing
the shape of the extensive low-energy ``red'' wing to the iron line.

\section{Analysis}

Fabian \et (2002a) discuss details of the observation and basic data
reduction. For the present paper the data were reprocessed entirely
with the latest software ({\tt SAS v5.3.3}) but the procedure followed
that discussed in the previous paper (except that this time the source
extraction radius was 35 arcsec). For both the EPIC pn and MOS data,
standard spectral redistribution matrices were used and ancillary
responses were generated using {\tt ARFGEN v1.48.10}. The source
spectra were fitted using {\tt XSPEC v11.1} (Arnaud 1996) and the
quoted errors on the derived model parameters correspond to a 90 per
cent confidence level for one interesting parameter (i.e. a $\Delta
\chi^{2}=2.7$ criterion), unless otherwise stated, and fit parameters
(specifically line and edge energies) are quoted for the rest frame of
the source.

\subsection{EPIC spectra}

\begin{figure}
\rotatebox{270}{
\resizebox{!}{\columnwidth}
{\includegraphics{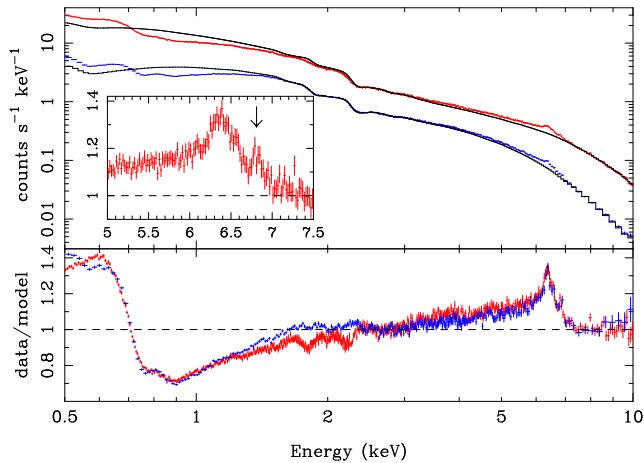}}}
\caption{
Top panel: EPIC pn (red) and combined MOS1+MOS2 (blue) spectra,
Bottom panel: Ratio of data to a power-law model joining the 2.5--3 keV data
and 7.5--10 keV data (cf. Fig.~1 of Fabian \et 2002a). 
As this is not a realistic model for the
continuum the residuals should be considered merely as representative
of the spectral complexity. The data from the two instruments obviously show 
slightly different spectral details. The inset panel shows the pn residuals
around the iron line region with the $\sim 6.9$~keV feature marked.
}
\end{figure}

Fig.~1 shows the time-averaged EPIC pn and MOS spectra for the entire
dataset compared to a simple power-law model fitted between
2.5--3.0~keV and 7.5--10.0~keV. The emission feature peaking at
6.4~keV is prominent, as is the absorption below 2~keV. However, as is
also clear from this plot, the detailed spectral features revealed by
the two detectors are subtly different. As the instruments were
operated simultaneously these differences represent differences in the
spectral calibration between the MOS and pn. In particular, the pn
shows a somewhat steeper spectral slope ($\Gamma = 1.97 \pm 0.08$ for
the pn compared to $\Gamma = 1.86 \pm 0.15$ for the MOS), a larger excess of
emission in the 3.5--5.5~keV band, and obvious sharp features around
$\sim 1.8$~keV and $\sim 2.2$~keV. These last two features are most likely associated with the
instrumental edges of Si-K at 1.84~keV and Au-M at 2.3~keV. The two
spectra also diverge strongly below $\sim 0.6$~keV.

Detailed spectral fits to the 2.5--10.0~keV spectrum were presented in
Fabian \et (2002a) based on the MOS data only. These gave a more
conservative estimate of the strength and breadth of the broad iron
line than the pn data. The greater excess in the pn data in the
3.5--5.5~keV band leads to an apparently stronger line with a
significantly stronger red wing. Fitting Model 3\footnote{Model 3
consists of a power-law continuum plus a {\tt pexrav} reflection
component ($R=1$) and an iron line, both relativistically blurred using
a broken power-law emissivity function.} of Fabian \et (2002a) to
the 2.5--10~keV pn spectrum gave a good fit ($\chi^2=1310.9$ for
$1492$ degrees of freedom, $dof$) with the following parameters:
continuum slope $\Gamma = 2.02 \pm 0.01$, disc inclination
$i=27.5\pm0.9$~deg, inner radius $r_{\rm in}=1.95\pm0.03 r_{\rm g}$,
break radius $r_{\rm br}=6.1_{-0.5}^{+0.8}r_{\rm g}$, inner emissivity
index $q_{\rm in}=5.5\pm0.3$, outer emissivity index $q_{\rm
out}=2.7\pm0.1$ and a line equivalent with of $EW=685_{-32}^{+50}$~eV.
The equivalent width of the possible $\sim 6.9$~keV emission line is
$EW=17\pm3$~eV.

It should be noted that while the above calibration problems remain
with the EPIC spectral responses, it is not clear whether the MOS or
the pn data give a more accurate description of the shape of the broad
iron line.  However, while the absolute values of best-fitting model
parameters are sensitive to detector-dependent calibration, the
spectral variability characteristics should be, to first order at
least, independent of these problems.  Therefore, the pn data are used
for the purposes of the spectral variability analysis below as they
have the highest signal-to-noise ratio. It should be noted that while
the form of the spectral variability should be
calibration-independent, the exact details of any spectral models will
remain sensitive to the calibration uncertainties outlined above.

\subsection{Spectral variability}

\begin{figure*}
\rotatebox{270}{
\resizebox{!}{2\columnwidth}
{\includegraphics{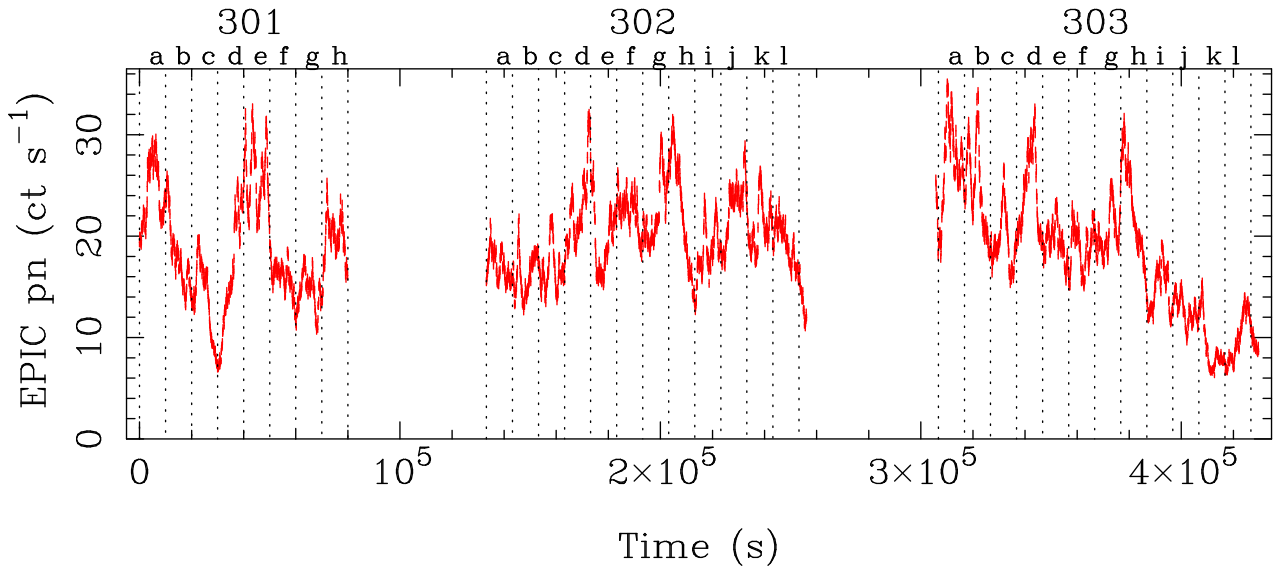}}}
\caption{
EPIC pn light curve (0.2--10~keV) showing the 10~ks intervals flux
discussed in the text. (This light curve has not been corrected for
the live-time of the pn camera in small-window mode, which is just a
scaling factor.) }
\end{figure*}

\subsubsection{The difference spectrum}

In this section the two-component
model discussed in Shih \et (2002) is applied to spectra obtained from
the \xmm\ observation. 
If the spectrum can be broken into two distinct emission components
then the observed spectrum during any given time interval, accounting for the
effects of absorption, can be written as  $S = A \times (N_1 s_1 + N_2 s_2)$, where
$s_1$ and $s_2$ represent the spectra of the two emission components,
$N_1$ and $N_2$ are their normalisations, respectively, and $A$ gives
the absorption profile. 

Shih \et (2002) found that such a model
adequately described the spectral variability observed in \mcg\ by
\asca, with one component having a constant flux while the other
varied only in normalisation. If the spectral variability can indeed be
described in terms of a variable emission component and another
component that remains constant (or, at least, varies very little on
the timescales of interest) then it becomes possible to examine the
absorption profile using the spectral variability. 
The difference between two spectra that differ
significantly in flux (different $N_1$ but 
identical $N_2$) is simply $A \times (N_{\rm 1, high} - N_{\rm
1, low})s_1$. The difference spectrum is thus the spectrum of the 
variable component only, modified by absorption. If the variable
spectral component is a power-law, as suggested by Shih \et (2002), then
the difference between two spectra of different fluxes should be a
simple power-law modified by (Galactic and intrinsic warm) absorption.

In order to examine this possibility the \xmm\ observation was divided
into $32\times 10$~ks intervals (Fig.~2) from each of which spectra
were obtained. The lowest-flux spectrum is from interval 303:k. A
high-flux spectrum was extracted from the first 80~ksec of data from
this revolution (303:a--h). The average count rate during the
high-flux interval is $32.6$~ct s$^{-1}$ and during the low-flux
interval the count rate is $13.5$~ct s$^{-1}$ (after correcting for
the live-time of the pn camera in small-window mode).  Subtracting the
two spectra (i.e. high$-$low), accounting for the difference in
exposure times, gives the difference spectrum shown in Fig.~3. 

\subsubsection{A simple absorption model}

Across the 3--10~keV band the difference spectrum is well fitted by a
power-law, with photon index $\Gamma = 2.11 \pm 0.05$. The strong,
broad iron line, while clear in both the high- and low-flux spectra
when examined separately, is not apparent in the difference spectrum.
The iron line flux must be very similar in these two spectra, even
though the total (i.e. continuum) flux differs by a factor of about 2
above 7~keV. This then implies a very strong line equivalent width
during the low-flux interval.

Extrapolating the power-law model to 0.5~keV reveals the attenuation
spectrum of the source (Fig.~3). The warm absorber is clear as the
large drop above $\sim 0.7$~keV which recovers by about $2$~keV, and
the Galactic absorption mostly accounts for the decrease at lower
energies. These features can be accounted for using a very simple
absorption model (similar to that used to model ASCA data of the
source; Otani et al 1996) comprising a power-law modified by
absorption from neutral gas of column density $N_{\rm H} = 3.4 \pm 0.4
\times 10^{20}$~cm$^{-2}$, explained by the Galactic column density
($4.06
\times 10^{20}$~cm$^{-2}$; Elvis, Lockman \& Wilkes 1989), together with two
warm absorption edges (at $E_1=0.72\pm0.005$~keV of depth $\tau_2 =
0.61\pm0.03$ and at $E_2 = 0.86\pm0.01$~keV with depth $\tau_2 =
0.19\pm0.04$). The obvious candidates for such edges are OVII and
OVIII, respectively.

\begin{figure}
\rotatebox{270}{
\resizebox{!}{\columnwidth}
{\includegraphics{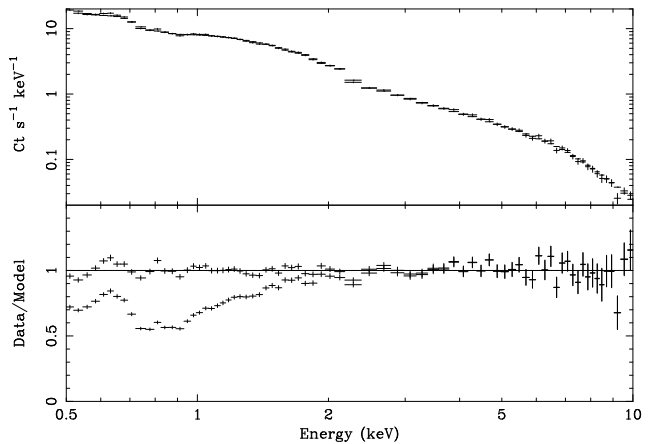}}}
\caption{
Top panel: The difference spectrum (high$-$low).
Bottom panel: Data/model residuals when the difference spectrum is
compared to a simple power-law model fitted only above 3~keV (red
crosses) and a power-law modified by a simple absorber (green circles).
}
\end{figure}

We note that the above simple absorption model may be adequate for
fitting data at CCD resolution above 1~keV but cannot be the whole
explanation for data below that energy, where absorption by lines and
Unresolved Transition Arrays are expected. The important point for the
present work is that there are no major absorption features associated
with iron-K, i.e. above 7~keV, nor with Si, S etc between 2--4~keV.
The broad iron emission line is not strongly affected by any
absorption features. Very similar results are obtained if the absolute
minimum, 11~ks, spectrum from within intervals 303:k and l is used.
Note that these results depend on the absorption not varying, which
is consistent with the rms variability spectrum (Fabian et al 2002a)
showing no edges.

\subsubsection{The low-flux spectrum}

The absorption model derived above was then applied to the spectrum
obtained from the low-flux interval. Fitting this spectrum with a
power-law, and including the absorption as described above, reveals
the very strong iron emission line and an excess of emission below
$\sim 1.5$~keV (Fig.~4). 

The iron line can be modelled by a {\tt LAOR} profile (Laor 1991) with
a high equivalent width ($EW=1.4$~keV) extending into an inner radius
of $1.8r_{\rm g}$ with an emissivity profile index of $3.3 \pm 0.3$.
The spectrum can also be fitted reasonably well with the constant
density ionised disc model of Ross, Young \& Fabian (1999), when
relativistic blurring is included (again using a broken emissivity
law). The model used included three times solar abundances and the
best-fitting ionisation parameter was $\xi = 10$, with the reflection
component dominating over the power-law ($R=5.5$). We hereafter refer
to this spectral model as the Reflection-Dominated Component (RDC) .

\begin{figure}
\rotatebox{270}{
\resizebox{!}{1\columnwidth}
{\includegraphics{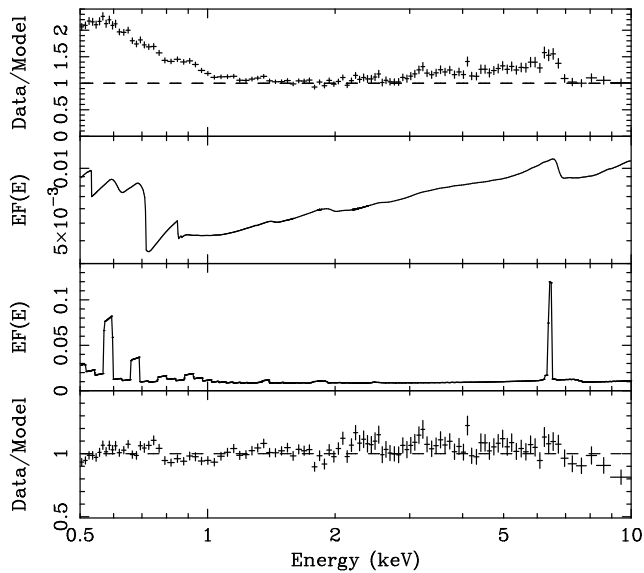}}}
\caption{From top to bottom the panels show: a) Data/model ratio plot
from fitting the lowest 10~ks flux spectrum with a power-law between
2--3~keV and 7.5--10.0 ~keV (including the effects of absorption as
derived from the difference spectrum); b) $EF_E$ model spectrum
including relativistic blurring and absorption effects; c) the
intrinsic model spectrum (note lines due to OVIII and OVII below
0.7~keV); and d) the data/model ratio of the best fitting model in the
1--10~keV band extended down to 0.5~keV. The drop in the ratio below
0.6~keV can be removed with the addition of another absorption
component.}
\end{figure}

\subsubsection{Fitting the whole light curve}

We now extend the study of the simple, absorbed, two-component model
to the whole dataset. A model consisting of the simple absorption
model from Section 2.2.2 applied to the best-fitting RDC and a
Power-Law Component (PLC),
with all parameters apart from the normalizations and PLC photon index
fixed, was then fitted to all 32 spectra over the energy
band 1--10~keV.

\begin{figure}
\rotatebox{270}{
\resizebox{!}{\columnwidth}
{\includegraphics{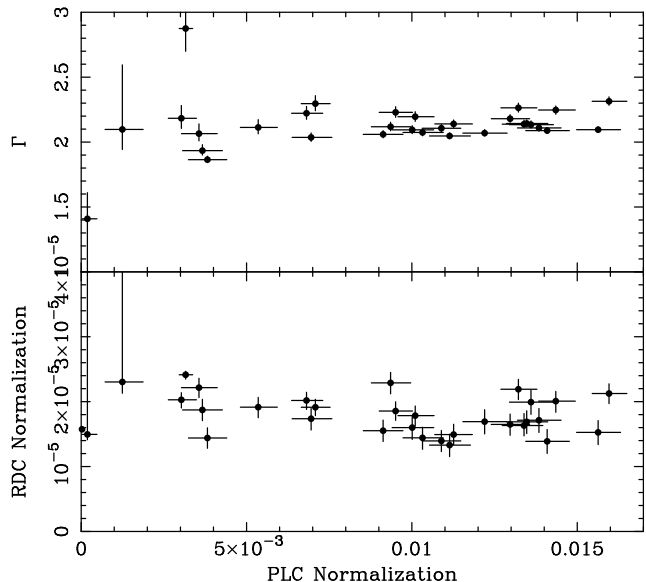}}}

\caption{
Top panel: Photon index $\Gamma$ of the PLC plotted versus the
normalization of the PLC. Bottom panel: Normalization of the RDC
plotted  against the PLC normalization.
}
\end{figure}

The results are shown in Fig.~5. All fits gave acceptable values of
$\chi^2$. The photon index $\Gamma$ of the PLC is seen to lie between
2 and 2.3 for most of the spectra. The deviant results occur when the
PLC normalization is low and disappear if the fit is made over the
2--10~keV band. It is possible that there are small changes in the
absorption when the flux is low. The RDC normalization is
approximately constant and shows variations of up to $\pm 25$ per
cent, with the variations showing no obvious correlation with other
parameters.

\section{Interpretation}

We have shown that the spectral variability of \mcg, on
timescales of 10~ks, is accounted for by an almost constant
reflection-dominated component together with a highly variable
power-law component, which has an almost constant photon index,
supporting the model of Shih et al (2001). We now attempt to interpret
the source using a two component RDC plus PLC model. An alternative
interpretation, in which the line is progressively ionized and
weakened as the source brightens, is presented and discussed by
Ballantyne, Vaughan \& Fabian (2002).

Explaining the constant emission component provides us with a
significant challenge. It appears to be mostly due to reflection.
However it is not simply due to reflection of the observed power-law
component since that repeatedly varies by factors of two or more on
short timescales (but not shorter than the inferred emission region
size, if it is indeed only a few $r_{\rm g}$). In other words the RDC
and PLC appear to be separate. Since however both show the effects of
the warm absorber they must originate in a similar location to within
the dimensions of the warm absorber (probably less than 1~pc). As the
iron line in the RDC indicates emission peaking at only a few
gravitational radii we shall henceforth assume that this is indeed
where that component originates. We then have to explain why so little
of the continuum which causes the line is observed and why the
power-law continuum we do see does not seem to have any associated
reflection.

We are driven to consider that the intrinsic X-ray emission is
anisotropic, so that the reflection component is a smoothed average of
many variable power-law components, most of which do not illuminate
our line of sight. This could be due to source geometry, relativistic
beaming of the emitters, or to the general relativistic bending of
light expected from an emission region so close to the black hole. We
now address each of these in turn.

\subsection{Geometrical effects}

As discussed by Fabian \et (2002b) in the case of 1H~0707--495, the
surface of the disc could be corrugated. If the continuum is emitted
close to the surface within the corrugations then it could show large
apparent variability as part of it disappears within a
corrugation. The reflected component could be strong, since we are
only seeing part of the intrinsic PLC (and multiple reflection may
operate; Ross \et 2002), and be approximately constant.

\subsection{Intrinsic beaming}

If the emitting region, or the emitting electrons, are relativistic,
then the emission can be beamed. We may see variations as beams from
different regions or populations of electrons sweep our line of
sight. Reflection from the disc however responds to the average of
many beams sweeping over it. Note that if the particles are
relativistic then the reflection can be strong due to the intrinsic
anisotropy of the inverse-Compton process (Ghisellini \et 1991).

\subsection{General relativistic effects}

If we accept the evidence from the line shape the the emission peaks
at say $2r_{\rm g}$, then the returning radiation (Cunninham 1975)
will be very important (see also Martocchia \et 2002; Dabrowski \et
1997). An observer on the disc at that radius would see much of the
Sky covered by disc, due to gravitational light bending. The
reflection is therefore strong and dominated by the average PLC
emission from the opposite side of the disc. The PLC which we see will
be dominated by the approaching side, perhaps associated with the
plunge region just within the marginally stable orbit (Krolik 1999,
Agol \& Krolik 2000). This will be a much smaller region, also Doppler
boosted, than that to which the reflection responds. If its height
varies with time then so will the flux of the PLC, whereas the RDC
will see a much more constant illumination and so vary little (see
Dabrowski \& Lasenby 2001, Fig.~12, for the expected variation in
equivalent width, dominated by changes in height of the PLC). The
appearance of the PLC and RDC are therefore essentially decoupled, as
required.

\section{Discussion}

The iron line in \mcg\ appears to indicate an extreme disc around a
spinning black hole. We have proposed several explanations for the
apparent constancy of the RDC and distinguishing between them is not
simple. General relativistic effects ought however to be important if
much of the emission originates at about $2r_{\rm g}$. They do provide
a simple explanation in which small changes in the height of the
power-law emission region lead to large changes in the observed flux
as a consequence of gravitational light bending; much of the flux is
bent onto the disk giving a strong, almost constant, reflection
component. This model will be pursued in detail elsewhere.

If instead the emission is anisotropic due to relativistic beaming,
then the PLC might extend as a power-law to high energies, perhaps
above 500~keV as seen in the hard tail in the soft state of Cygnus
X-1. 

In summary, we find that the spectrum of \mcg\ is well
represented by the sum of an almost constant reflection-dominated
component and a highly variable power-law component. We emphasise that
this is only an approximation to the spectral variability as it does
not account for the observed changes in the power-spectral density
with energy, nor the hard lags (Vaughan et al 2002). Acting on both is
a warm absorber, which above 1~keV is relatively simple and dominated
by highly-ionized oxygen. Provided that the emission predominantly
originates from about $2r_{\rm g}$, then the strengths of the
components and their behaviour are accounted from by strong
gravtiational light-bending. The black hole in this case is presumably
spinning rapidly.

Strong broad iron lines in active galactic nuclei may require the
central black hole to have a high spin in order that returning
radiation creates a strong reflection component. If the rapid
variability in \mcg\ is in part due to strong light-bending
amplifying small changes in source height, then the similarity with
the timing behaviour of the soft state of Cygnus X-1 (Vaughan, Fabian
\& Nandra 2002) suggests that it too may be rapidly spinning. Further
extrapolation to the variability of many Narrow-Line Seyfert 1 galaxies 
(e.g. Boller et al 1997) is also interesting.

\section*{Acknowledgements}
Based on observations obtained with \xmm, an ESA science mission with
instruments and contributions directly funded by ESA Member States and
the USA (NASA). We thank David Ballantyne, Russell Goyder, Kazushi
Iwasawa, Anthony Lasenby and Andy Young for discussions. ACF thanks
the Royal Society for support.

\bsp
\label{lastpage}
\end{document}